# A probabilistic model for pedestrian gap acceptance behavior at uncontrolled midblock crosswalks

**Ali Mansour Khaki[1*], Amin Mohammadnazar[2]**

[1] Associated Professor, Department of Civil Engineering, Iran University of Science & Technology, Tehran, Iran
[2] Department of Civil Engineering, Iran University of Science & Technology, Tehran, Iran

**Abstract**

A significant number of those killed in traffic accidents annually refers to pedestrian. most of these accidents occur when pedestrian is going to pass the street. One of the most hazardous areas for pedestrian crossing the street is midblock crosswalks. These areas are often uncontrolled and there is no separated phase for pedestrian crossing. As a result, using created gaps among vehicles is often the only opportunity for pedestrian to cross the street. This research aims to investigate effective factors on pedestrian gap acceptance at uncontrolled midblock crosswalks. In this regard, several variables such as individual, environmental and traffic variables were considered and their related data were extracted by filming the location. from video images, they were analyzed using logit regression model and the behavior After extracting the data of pedestrian while he is crossing the street was modeled as the probabilistic function of pedestrian gap acceptance. The results illustrate that conflict flow rate, temporal gap among vehicles and driver yielding in the face of vehicles are of the most important effective factors on the pedestrian behavior. The other one is that, the speed and type of vehicles don't affect pedestrian decision about rejecting or accepting the gap. The results also show that individual characteristics such as age and gender of pedestrian don't make significant relationship with pedestrian gap acceptance.

[*] Corresponding author.Tel.: +98 21 7454056 Ex: 3144.
fax: +98 21 77454053
E-mail address: mkhaki@iust.ac.ir

## 1- Introduction

One of solutions for ameliorating pedestrian security is creating safe routs for them. Although it is tried not to conflict with special equipment for the passage of vehicles, as these routes are designed and executed, most of the times, the conflict of pedestrian with vehicles will be inevitable. In these conditions, appropriate management of conflicting between pedestrians and vehicles is a vital issue. one of the commonplace solutions in this case is using traffic light and providing separated phase for pedestrians to cross; yet using traffic light, in areas which the passage of pedestrians and vehicles isn't noticeable, isn't justifiable and if it is justifiable, some of considerations prevent the installation of traffic lights (MUTCD). Consequently, pedestrian crossing is conducted uncontrolled in most of the areas. Most of the times there is no traffic light in pedestrian midblock crosswalks and as a result, identifying pedestrian right of way while crossing midblock pedestrian crosswalk is very difficult. Although traffic laws in some countries gives the right of way to the pedestrians in this case, field observations show that not all drivers follow this rule. Whenever drivers don't give the right of way to the pedestrians, the pedestrians have to use the gap among vehicles to cross. In other words, most of the times, using created safe gap among crossing vehicles is the only opportunity of pedestrian to cross the street. This case has turned midblock crosswalks to one of the most dangerous areas for pedestrian crossing. According to the report of The World Health Organization in 2015, 251 million people die in the road accidents annually in which pedestrians are involved in 22 percent of them. Most of these accidents occur when the pedestrians are going to cross the street (WHO 2015). In Iran also according to the report of legal medicine organization in 2015, 16 thousands and 584 hundreds of people have been killed in driving accidents in which 22.2 percent that is 3 thousands and 688 hundred of were pedestrians (IMO 2015).

What makes the importance of investigating pedestrians' behavior in midblock crosswalks is double higher crash rate in these areas than the other ones. Through assessing 786 samples of children crash with vehicles which had happened a one year range in Montreal, Canada, David and Rise (1994) showed that most of these accidents (63%) had happened in midblock crosswalks and urban areas. In another work, Cho et al (2006) conducted some researches about the security of pedestrians in midblock areas in Florida, America. Investigating pedestrian accident death toll showed that accident death toll has been significantly increased in midblock areas so that within 1994-2001, 81% of whole accidents leading to death and 3.77% of them leading to injury had happened in midblock areas. The toll of fatal accidents shows that crossing the streets in midblock areas is more fatal than crossing the intersections. For example, the percentage of fatal accidents to total accidents in midblock was 2.8% and for intersections 6.5%.

The objective of this study is to to evaluate effective factors on pedestrian gap acceptance in uncontrolled midblock crosswalk.

## 2- Literature review

Pedestrians' behavior while crossing the street has been studied in many researches. The age and gender of pedestrians were among the characteristics that have been always studied. Hamed (2001) figured out that waiting time for pedestrians before crossing the street is in direct relationship with pedestrians' age. Male pedestrians also wait longer than female ones. The studies also show that older pedestrians choose longer gaps for crossing the street than younger ones (Oxley et al (2005 and 1997) and Lobjois and Cavallo (2007)). The results of researches by Connelly et al (1998 and 1996) showed that children are weak in identifying and assessing the speed of vehicles and only make decision according to the gap of the closest vehicle.

Pedestrian gap acceptance has been investigated in many studies. Pedestrian gap acceptance studies are generally divided into two critical and probabilistic categories. In critical ones, final output of most of studies is proposing pedestrian gap acceptance as critical gap. By definition, critical gap is time gap between successive vehicles in the main flow, the one which pedestrian may accept or reject this gap with equal probability for crossing the street. HCM, 2010 has proposed a method for estimating pedestrian critical gap acceptance in which accepted gap is as a function of crosswalk length, speed of pedestrian and the time of beginning the movement. Rouphail et al (2005) have described pedestrian gap acceptance as the sum of delay and real time of passage and used field studies for estimating the time of delay. In another research in China, Yung et al (2006) considered critical gap as the function of crosswalk length, speed of pedestrian and pedestrian confidence while crossing the street. The studies show that critical models don't have that much accurate estimation of pedestrian gap acceptance and most of obtained values from them are much larger than the actual values. On the contrary, using probabilistic models for assessing the behavior of pedestrians while crossing the street have brought more accurate results. Also among probabilistic models, logit regression models have showed better fit with data than neural models and linear regression (Sun et al, 2002; Brewer et al, 2006; Kadali et al, 2015 and Pawar and Patil, 2016).

In probabilistic studies unlike critical ones, pedestrian's behavior is different under different situation. Many studies have been conducted over the recent years in this case in which the effect of individual and environmental characteristics on the behavior of pedestrians has been investigated. These studies show that by increasing the gap between vehicles, pedestrian gap acceptance will be more likely (Sun et al, 2002; Oxley et al, 2005; Brewer et al, 2006; Cherry et al, 2012; Kadali and Vedagiri, 2013; Yannis et al, 2013). Group movement of pedestrians has been another controversial subjects in gap acceptance studies. In some of these studies group movement of pedestrians has led to longer gap acceptance (Yannis et al, 2013; Pawar and Patil, 2015) and in some other ones it has led to shorter gap acceptance (Hymanen et al, 1988). These studies also show that by increasing waiting time, pedestrian gap acceptance probability will decrease and pedestrians choose relatively longer gaps for crossing (Sun et al, 2002; Yannis et al, 2013; Cherry et al, 2012). Brewer et al (2006) figured out that for crossing street, instead of waiting till depletion of all lanes, the pedestrian accept first gap in first lane through predicting that while crossing the first lane, safe gap will be provided in the second lane. The authors of this paper have called this gap as rolling gap. The results of Kadali and Vedagiri studies in 2012 and 2013 showed that using or not using rolling gap has the most effect on the size of gaps accepted by pedestrians. Through investigating pedestrian jaywalk in China streets, Cherry et al (2012) figured out that probabilistic pedestrian gap acceptance has negative relationship with the speed of vehicles. The studies if Yannis et al, 2013 showed that illegal parking vehicles in crosswalks causes that pedestrian use longer gaps for crossing the street. Pawar and Patil (2014) showed that whatever the vehicle is bigger, pedestrian gaps acceptance are larger.

The studies above show that in conducted probabilistic studies about pedestrian gap acceptance, the role of flow's characteristics in pedestrian's behavior has been investigated less. On the contrary, in most of conducted studies, individual and environmental characteristics have been more paid attention. In this study, in addition to individual and environmental characteristics, the effect of characteristics of traffic flow on pedestrian's behavior and his gap acceptance have been studied.

### 3- Methodology

Main steps of this research includes choosing field study location, extracting data and developing pedestrian gap acceptance model. Each one of these steps is going to be discussed as following.

### 3-1- Choosing the location of field study

A crosswalk located in Tehran, Iran was selected as the location of field study. This crosswalk is midblock and uncontrolled type and located on three lanes to the width of 11 meters and has striping for pedestrian crossing. The street functions as one way for buses and the other one for other vehicles. This part of street has been separated by metal fences in the middle of street so that pedestrians, after crossing bus-only lanes, can only cross the street using crosswalk. Figure 1. shows the location of crosswalk in detail.

### 3-2- Data collection and extraction

The location was filmed using a digital camera which was installed in the site of a commercial building overlooking the crosswalk. Filming was conducted for 90 minutes from 4:30 to 6 p.m. and on a weekday. Extracted video images covered simultaneously crosswalk and 35 meters of street length.

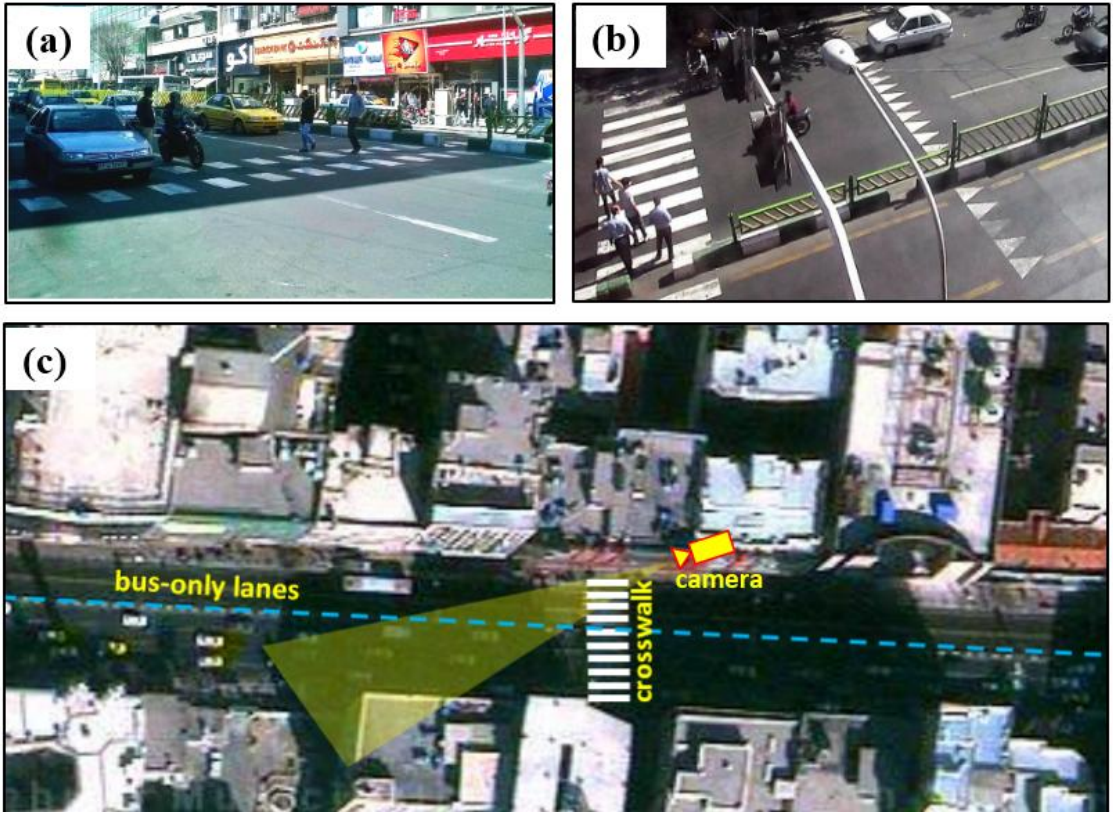

Figure 1- (a) aerial photo, (b) camera view, (c) street view

The behavior of 277 pedestrians while crossing the street was extracted during filming in which the result was 1359 conflicts between pedestrian and vehicle. Among this, 760 gaps were accepted by pedestrians and 599 gaps were rejected. considering the characteristics of flow, conflict flow rate was used. Conflict flow rate was first defined by Kite et al in 1991 and for assessing vehicle gap acceptance. Chandra et al (2013) used conflict flow rate for assessing pedestrian critical gap acceptance. Conflict flow rate is proposed as equation 1.

$$Conflict\ flow\ rate = \frac{n}{t_2 - t_1} \qquad (1$$

In which:

n= the number of vehicles that are conflicted by pedestrian from the moment he reaches to the crosswalk till the time he decides to cross the street.
$t_1$= Arrival time of pedestrian in the site of crosswalk
$t_2$= The time of last vehicle passes through the site of crosswalk

in order to extract data such as the speed of pedestrian and vehicles, *Tracker video analysis* program was used. Table 1 shows the list of potential variables with specific values to each one. Among the available variables in this table, variable ACC is dependent one and others are independent variables.

Table 1- Effective potential variables on pedestrian gap acceptance

| Variable | Type | Definition | Unit |
|---|---|---|---|
| ACC | Discrete | whether pedestrian accepts created gap for crossing or not | 0= no; 1=yes |
| GEN | Discrete | Gender of pedestrian | 0= female; 1=male |
| GR | Discrete | whether pedestrian moves in group or alone | 0= alone; 1=in group |
| AGE | Discrete | The age of pedestrian through visual judgement | 0=young; 1=middle-aged; 2= old |
| DIR | Discrete | The direction of pedestrian movement from metal fences toward sidewalk or vice versa | 0=sidewalk; 1=fences |
| ROLL | Discrete | Use or non-use of rolling gaps by pedestrian | 0= non-use; 1=use |
| RUN | Discrete | whether pedestrian starts running while he is crossing the street or not | 0= no; 1=yes |
| PARK | Discrete | The presence of parked vehicles in crosswalk illegally | 0= no; 1=yes |
| VEH | Discrete | The type of vehicle (motorcycle, car, heavy vehicles) | 0=motorcycle; 1=car; 2=heavy vehicles |
| LAG | Discrete | whether pedestrian use lag or gap for crossing * | 0= non-use; 1=use |
| YIELD | Discrete | Driver yielding | 0= no; 1=yes |
| LANE | Discrete | The number of lane in which pedestrian conflicts with vehicle | 1=near lane; 2=far lane; 3= farther one |
| LOAD | Discrete | whether the pedestrian carry a load or not | 0= no; 1=yes |
| CELL | Discrete | Whether the pedestrian is talking on the phone | 0= no; 1=yes |
| PS | Continues | The speed of pedestrian | m/s |
| RATE | Continues | Conflict flow rate | veh/s/ln |
| TIME | Continues | Waiting time of pedestrian before crossing | s |
| VEHS | Continues | The speed of vehicle | m/s |
| GAP | Continues | Temporal gap between vehicles | s |

* When the pedestrian reaches crosswalk, the first created gap between the first vehicle and pedestrian is so-called LAG and next gaps which happen between two vehicles is so-called GAP.

Table 2 shows descriptive statistics of variables which are extracted from video files. According to this table, the mean 0.73 for variable GEN shows that 73% of observed pedestrians were men and the rest were women. Also the mean of GR shows that 52% of pedestrians have crossed the street in group. The percentage of driver yielding was only 20%. The mean of variable ROLL represents the rate of using rolling gaps by pedestrians. The value of mean in this variable shows that 82% of pedestrians use rolling gaps to cross the street which is significant. The mean of waiting time for pedestrians before crossing (TIME) is 8.35s.

Table 2- Sample descriptive statistics

| Variable | Minimum | Maximum | Mean | Standard deviation |
|---|---|---|---|---|
| Discrete | | | | |
| ACC | 0.00 | 1.00 | 0.56 | 0.50 |
| GEN | 0.00 | 1.00 | 0.73 | 0.44 |
| DIR | 0.00 | 1.00 | 0.63 | 0.48 |
| ROLL | 0.00 | 1.00 | 0.83 | 0.38 |
| LOAD | 0.00 | 1.00 | 0.13 | 0.34 |
| RUN | 0.00 | 1.00 | 0.04 | 0.20 |
| PARK | 0.00 | 1.00 | 0.23 | 0.42 |
| CELL | 0.00 | 1.00 | 0.03 | 0.17 |
| VEH | 0.00 | 3.00 | 0.73 | 0.50 |
| LAG | 0.00 | 1.00 | 0.51 | 0.50 |
| YIELD | 0.00 | 1.00 | 0.20 | 0.40 |
| LANE | 1.00 | 3.00 | 1.66 | 0.80 |
| GR | 0.00 | 1.00 | 0.52 | 0.50 |
| AGE | 0.00 | 2.00 | 0.54 | 0.74 |
| Continuous | | | | |
| PS | 0.84 | 2.90 | 1.37 | 0.32 |
| TIME | 0.00 | 56.00 | 5.29 | 11.27 |
| VEHS | 1.10 | 15.80 | 7.03 | 2.47 |
| RATE | 0.077 | 1.52 | 0.462 | 0.20 |
| GAP | 0.32 | 22.00 | 3.29 | 2.33 |

### 3-3- Developing pedestrian gap acceptance model

Statistical software SAS9.4 was used in order to develop pedestrian gap acceptance model. Full model of pedestrian gap acceptance with all potential variables was first developed through logit regression method. Wald test was utilized in order to assess the significance of independent variables and confidence level was considered as 95%. On next step, those variables which couldn't make a significant relationship with response variable (p-

value<0.05) were eliminated and remained variables were proposed in form of restricted model. Table 3 shows the results of analyzing full model and restricted model.

Table 3- analysis results of full and restricted logit model of pedestrian gap acceptance

| Parameter | Coefficient | Standard error | Wald | p-value |
|---|---|---|---|---|
| Full model | | | | |
| Intercept | -13.737 | 1.441 | 90.849 | <.0001 |
| PS | 1.868 | 0.526 | 12.621 | 0.000 |
| GEN | -0.100 | 0.308 | 0.106 | 0.745 |
| GR | -0.614 | 0.288 | 4.561 | 0.033 |
| AGE | -0.032 | 0.225 | 0.020 | 0.887 |
| TIME | -0.089 | 0.017 | 26.464 | <.0001 |
| DIR | 0.179 | 0.333 | 0.291 | 0.590 |
| ROLL | 0.832 | 0.382 | 4.739 | 0.030 |
| LOAD | -0.095 | 0.409 | 0.054 | 0.817 |
| RUN | -0.315 | 0.594 | 0.281 | 0.596 |
| PARK | -0.236 | 0.371 | 0.405 | 0.524 |
| CELL | 0.298 | 0.863 | 0.119 | 0.730 |
| VEH | -0.318 | 0.305 | 1.087 | 0.297 |
| GAP | 1.999 | 0.153 | 170.799 | <.0001 |
| VEHS | -0.012 | 0.062 | 0.036 | 0.850 |
| LAG | 1.075 | 0.305 | 12.398 | 0.000 |
| YIELD | 1.909 | 0.359 | 28.271 | <.0001 |
| RATE | 4.511 | 0.890 | 25.719 | <.0001 |
| LANE | 1.663 | 0.219 | 57.560 | <.0001 |
| Restricted model | | | | |
| Intercept | -14.153 | 1.273 | 123.564 | <.0001 |
| PS | 1.817 | 0.472 | 14.847 | 0.0001 |
| GR | -0.604 | 0.259 | 5.431 | 0.020 |
| TIME | -0.089 | 0.015 | 35.715 | <.0001 |
| ROLL | 0.797 | 0.368 | 4.690 | 0.030 |
| GAP | 2.021 | 0.152 | 176.596 | <.0001 |
| LAG | 1.158 | 0.293 | 15.643 | <.0001 |
| RATE | 4.846 | 0.868 | 31.156 | <.0001 |
| YIELD | 1.854 | 0.341 | 29.530 | <.0001 |
| LANE | 1.593 | 0.208 | 58.401 | <.0001 |

in order to assess goodness of fit in models, *Bayesian information criterion (BIC), Akaike information criterion (AIC)* and *-2 Log Likelihood* were used. Lower values of BIC and AIC statistics represent better fit of logit regression model. Table 4 shows the values of these statistics for two full and restricted model. According to this table, restricted model has shown better fit with data.

Table 4- comparison of models' goodness of fit

| Statistics | Full model | Restricted model |
|---|---|---|
| AIC | 462.051 | 448.382 |
| BIC | 555.656 | 496.648 |
| -2 Log L | 424.051 | 428.382 |

Ultimately, final logit regression model of pedestrian gap acceptance was proposed by pedestrians as equation 2 and equation 3.

$$U' = -14.153 + 1.817\ PS - 0.604\ GR - 0.089\ TIME + 0.797\ ROLL + 2.021\ GAP + 1.158\ LAG + 1.854\ YIELD + 1.593\ LANE + 4.846\ RATE \quad (2)$$

$$P(Y = 1) = \frac{1}{1 + e^{-U'}} \quad (3)$$

In which:
$U'$= the index of pedestrian gap acceptance desirability
P= pedestrian gap acceptance probability in order to verifying the model, remained one-fourth of data, which were not used in modeling, were evaluated. The results showed that model has had a correct prediction of pedestrian behavior in 92.35% of occasions. (table 5) shows the verification results of gap acceptance model with separation of rejected and accepted gaps.

Table 5- the verification results of gap acceptance model with separation of rejected and accepted gaps

| | Accepted gaps | | Rejected gaps | |
|---|---|---|---|---|
| | Accepted in the model | Rejected in the model | Accepted in the model | Rejected in the model |
| proportion | 52.05% | 2.05% | 5.3% | 40.6% |

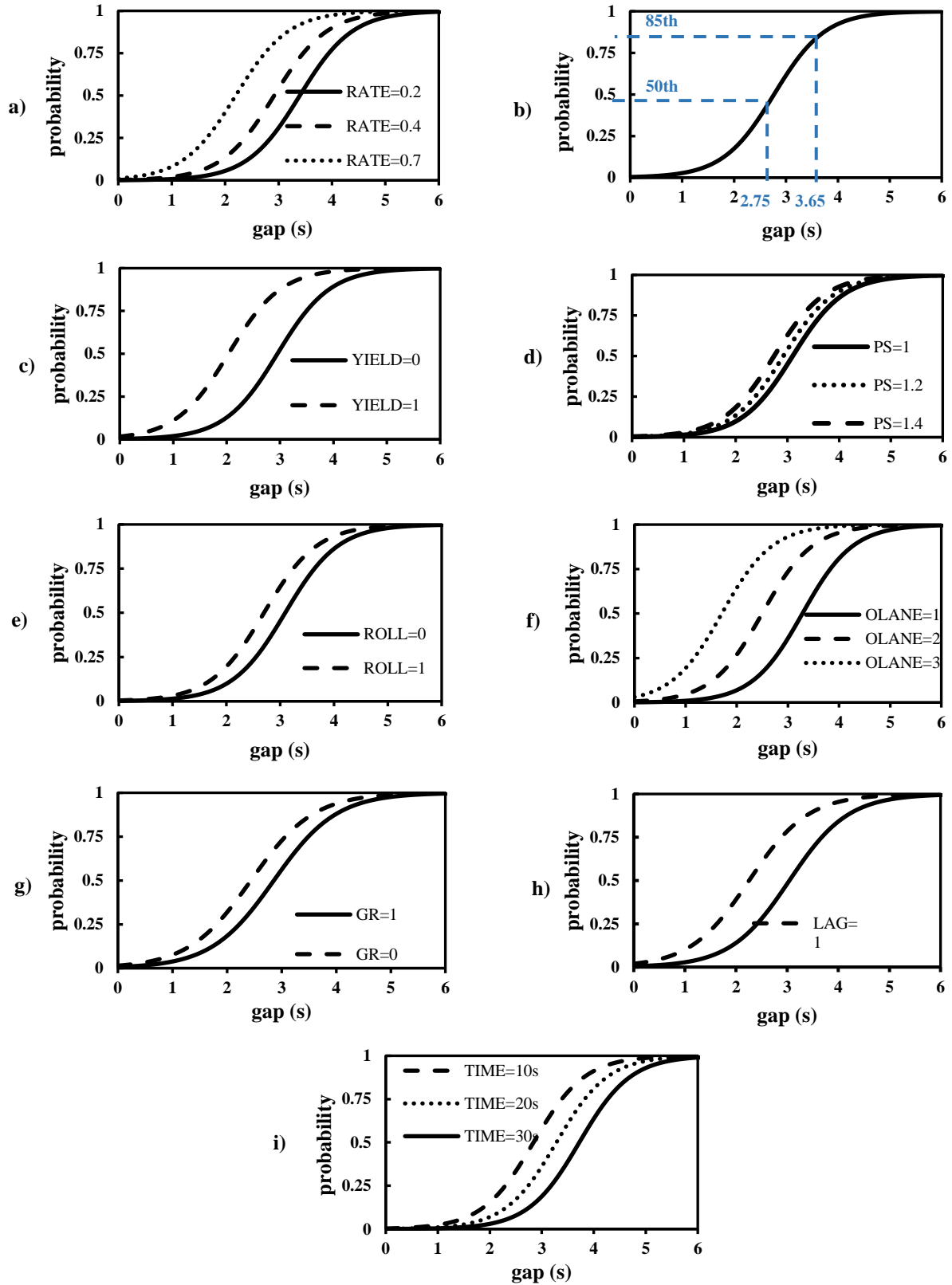

Figure 2- the diagram of pedestrian gap acceptance probability sensitivity to independent variables

1- Model results

Pedestrian gap acceptance model has 9 independent variable including GAP, ROLL, TIME, GR, PS, RATE, YIELD, LAG and LANE in which the variables ROLL, TIME, PS, GR and LAG are related variables to the characteristics of pedestrians and variables YIELD, GAP, RATE and LANE are related to environmental and traffic characteristics.

This shows that pedestrian, environmental and traffic characteristics are all effective on how pedestrians behave. Among these variables, the variables YIELD, GAP, RATE and PS have respectively got the largest coefficients and odds ratios inside model. The speed and type of vehicle didn't make a significant relationship with pedestrian's behavior in this study. In case of vehicle speed, recognizing the speed of vehicle seems to be difficult for pedestrian because of conflicting pedestrian and vehicles face to face. Individual characteristics such as age, gender of pedestrians couldn't also define response variable accurately. In fact, if all conditions are equal to pedestrians, pedestrian gap acceptance probability, whether male or female, is equal.

In order to investigate the effect of each one of variables on pedestrian gap acceptance, sensitivity analysis was used. Figure. 2 shows the diagram of sensitivity for each one of variables to the size of gap. The effect of each one of parameters involved in pedestrian gap acceptance will be investigated as following.

### 4-1- The impact of conflict flow rate

Conflict flow rate is very effective on determining the behavior of pedestrian. Positive sign of variable RATE in gap acceptance model represents that by increasing conflict flow rate, probabilistic pedestrian gap acceptance will increase. Figure 2-a shows sensitivity of gap acceptance probability to traffic gaps for different conflict flow rates. According to this diagram, for average gaps such as 3s gap, gap acceptance probability in small rates (200 veh/In/s), average (500 veh/In/) and large (700 veh/In/) will be respectively 31%, 54% and 84%. Figure. 3 shows the dispersion of flow rate to the size of accepted gaps. According to the figure, by increasing flow rate, pedestrians have chosen smaller gaps to cross the streets. In fact, the pedestrians who conflict with high rate of flow, if aren't interested in long waiting, have to accept smaller gaps to cross the street.

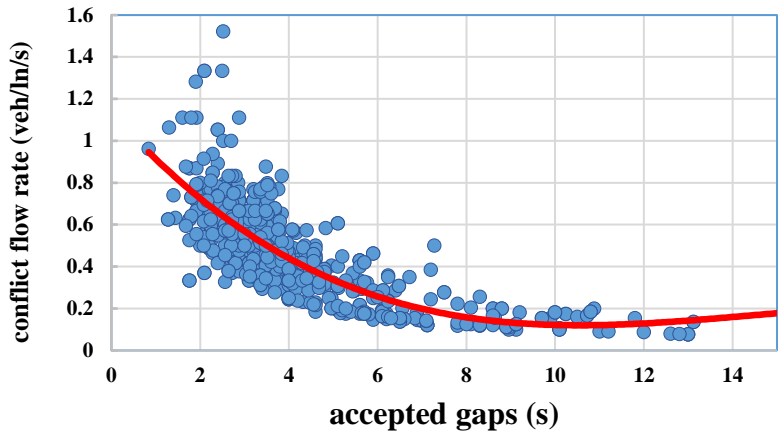

figure 3- the dispersion of flow rate to the size of accepted gaps by pedestrians

## 4-2- The impact of gap size

Positive sign of variable GAP in pedestrian gap acceptance represents direct relationship of gap size with the probability of gap acceptance. By increasing the distance between vehicles, the pedestrians will tend more to cross the streets. Figure. 2-b shows the diagram of gap acceptance probability sensitivity by pedestrian to the size of gap (GAP). According to this diagram, by increasing the size of gap, gap acceptance probability will increase by pedestrians so that the bigger gaps than 6 seconds will be accepted by all pedestrians (p=1). As an example, pedestrian gap acceptance probability in gaps 2s and 3s are respectively 60% and 93%. These values show that by increasing gap from 2s to 3s, pedestrian gap acceptance probability will increase by 33%. Gap of 2.75s corresponding to the 50th percentile shows accepted critical gap by pedestrian. In another word, a half of pedestrian which face the gap of 2.75s, accept it and the other half reject it. $85^{th}$ percentile on diagram also shows that 85% of pedestrians will accept the gaps of 3.65s. in case the speed of vehicle is assumed equal to the mean of vehicle speed in this research as 7m/s, spatial gap corresponding to 50 and $85^{th}$ percentile will be respectively equal to 19.25m and 25.55m.

## 4-3- The impact of driver yield

Variable YIELD as an index of drivers' behavior in conflicting with pedestrians plays an important role in how pedestrians behave while crossing the street. According to table 2, the value of odds ratio for YIELD variable is equal to 6.38 which is relatively significant. In another word, in case that the drivers of vehicles, as approaching to crosswalk, decide to help pedestrian through slowing down or changing lane, this issue will make the probability of gap acceptance to that acceptance rejection 6.38 times more. Figure. 2-c shows the sensitivity of pedestrian gap acceptance to the size of gap for different values of YIELD variable. According to the diagram, by drivers yielding for crossing the street, pedestrian gap acceptance probability will increase. Although this increase isn't that much big in small or big gaps but whatever it goes to average gaps, the difference will be more. For example, in facing a pedestrian with 2.5s gap, if the driver decides to slow down or change lane, pedestrian gap acceptance probability will increase by 44%. Based on this, figure. 4 shows the diagram of pedestrian accepted gaps dispersion in two modes of YIELD=0 and YIELD=1. According to this diagram, the pedestrians who have been let to cross the street by the drivers of vehicles, have accepted relatively shorter gaps for their crossing.

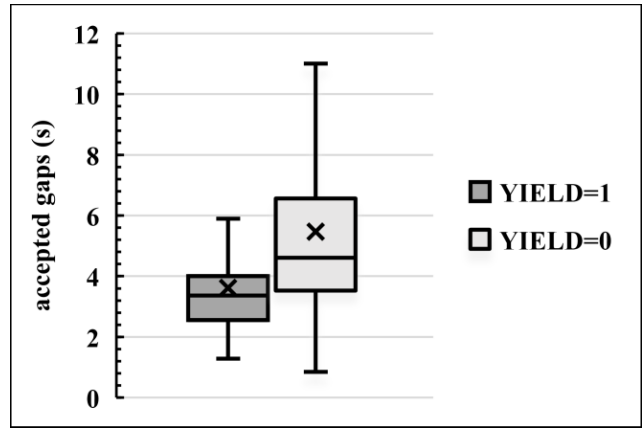

Figure 4- pedestrian gap acceptance for YIELD=0 and YIELD=1

### 4-4- The impact of pedestrian speed

The speed of pedestrian has a direct relationship with pedestrian gap acceptance probability. Figure. 2-d shows the sensitivity of pedestrian gap acceptance probability to the size of gap (GAP) for the speeds of 1.0 m/s, 1.2 m/s and 1.4 m/s. according to this diagram for a fixed value of probability, the pedestrians who choose shorter gaps for crossing the streets, cross the street faster. This shows that pedestrians usually adapt their speed with the size of gap to have safer crossing. In another word till created gap between vehicles is big enough, the pedestrians don't feel any need to increase their speed while crossing the street. But in case the pedestrian has to use shorter gaps because of different reasons, he increases his speed inevitably as far as cases that alarmed pedestrian start running.

### 4-5- The impact of rolling movements of pedestrians

The results illustrate that the probability of gap acceptance by pedestrians who use rolling gaps to cross the street is more. The investigation of extracted data from pedestrians show that almost 82% of pedestrians have used rolling gaps to cross the street which is significant. Figure. 5 shows the dispersion of flow rate for two modes of use or non-use of rolling gaps by pedestrian. This figure illuminates that rolling behavior often happens in high values of flow rate. Figure. 2-e shows the diagram of gap acceptance sensitivity diagram compared to the size of gap and Figure. 6 shows dispersion of gaps accepted by pedestrians who have or haven't used rolling gaps. According to these diagrams, the pedestrians who have used rolling gaps, have accepted relatively shorter gaps and, as a result, have more dangerous behavior.

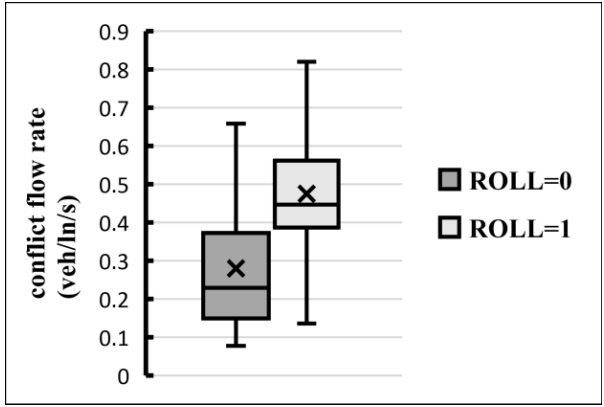

Figure 5- dispersion of flow rate for use or non-use of rolling gaps by pedestrian

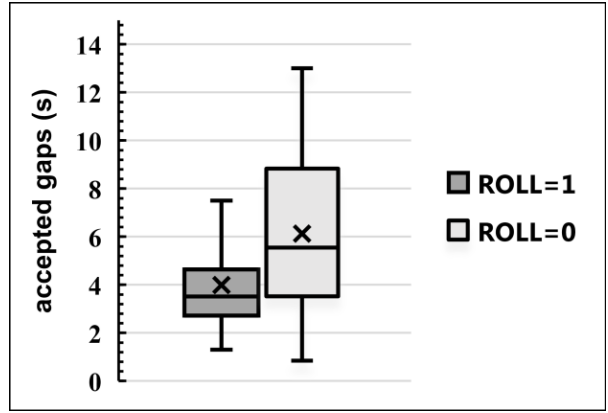

Figure 6- the dispersion of accepted gaps use or non-use of rolling gaps by pedestrian

## 4-6- The impact of conflict status

LANE variable which introduces the number of conflict lane, has 3 various levels. Whatever the value of this variable gets more, it shows how much the pedestrian gets far from the beginning point. Positive coefficient of this variable in model shows that whatever the pedestrian gets far from his initial location, his gap acceptance probability will be more. In another word, after accepting the first gap and crossing the first lane, when they reach to the middle of the street, the pedestrians inevitably have to choose shorter and more dangerous gaps to be able to reach the safe place (sidewalk) as fast as possible. Figure. 2-f shows the diagram of pedestrian gap acceptance probability sensitivity compared to the size of gap (GAP) for different values of variable LANE. According to this figure for gap of 2s which is considered as relatively small gap, the probability of gap acceptance in first lane is about 7%. This is while, by getting far from the initial point, pedestrian gap acceptance in the second and third lanes have increased significantly and respectively reaches 27% and 64%. Figure. 7 shows cylindrical diagram of accepted gaps in different lanes. This diagram well shows that how the pedestrians accepts shorter gaps in second and third lanes compared to the first lane.

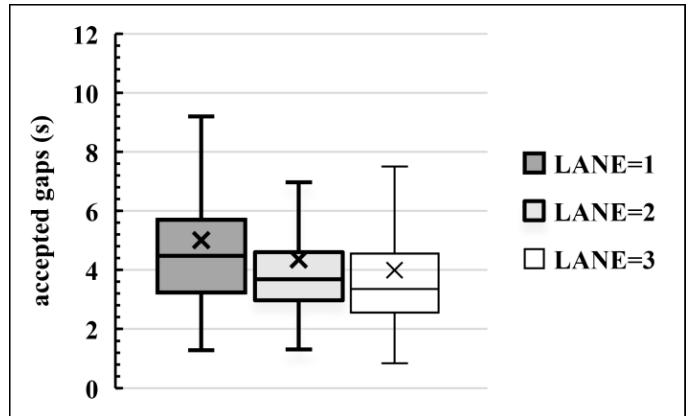
Figure 7- pedestrian accepted gap in different lanes

### 4-7- The impact of group movement of pedestrians

When the pedestrians cross the street in group, they usually have a similar behavior and often accept equal gaps. Negative sign of variable GR in pedestrian gap acceptance model represents that the pedestrian's movement in group will reduce the probability of gap acceptance. In another word, the pedestrians who cross the street in a group are more cautious than those who cross alone. Figure. 2-g shows the diagram of pedestrian gap acceptance probability sensitivity compared to the size of gap for pedestrian with group or lonely movement. According to the figure, in very short or very large gaps, the pedestrian group movement doesn't affect pedestrian gap acceptance probability that much and the gaps which have been accepted by pedestrians in group as alone are almost equal. In average gaps, this difference is more considerable so that for gap 3s, gap acceptance probability for group movement is 14% more than moving alone. Figure. 8 shows the pedestrians accepted gaps dispersion in group or lonely movement. This figure illustrates that the pedestrians who move in groups choose longer gaps than those who move alone.

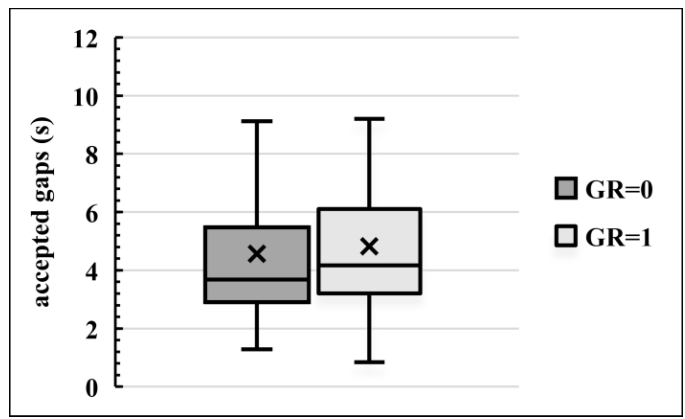
Figure 8- pedestrian accepted gaps for group or lonely movement

## 4-8- The impact of gap type

Variable LAG introduces the type of accepted gap by pedestrian. Positive sign of variable LAG in model shows that if created gap is LAG type, compared to when it is GAP type, pedestrian gap acceptance probability is more. Figure. 2-h shows the diagram of pedestrian gap acceptance probability sensitivity compared to the size of gap for different gap types. According to this diagram the value of accepting gap of 3s by pedestrians for when the gap is LAG type is 27% more than other gaps. This can be because the pedestrians who use the very initial gap to cross the street are more risk taking than those who waits for later gaps and as a result their gap acceptance probability is more. Investigating the behavior of pedestrians separating by lanes shows that the pedestrians tend more to accept the first gap in farther lanes. Figure. 9 represents the percentage of pedestrians who choose the first gap for crossing by separating the lane. According to this diagram whatever the pedestrian gets farther from his initial point, choosing the first gap for crossing street will be more likely. In another word, whatever the pedestrian gets farther from initial point, he feels less secure and as a result tries to get the safe place as fast as possible through accepting the first initial gap.

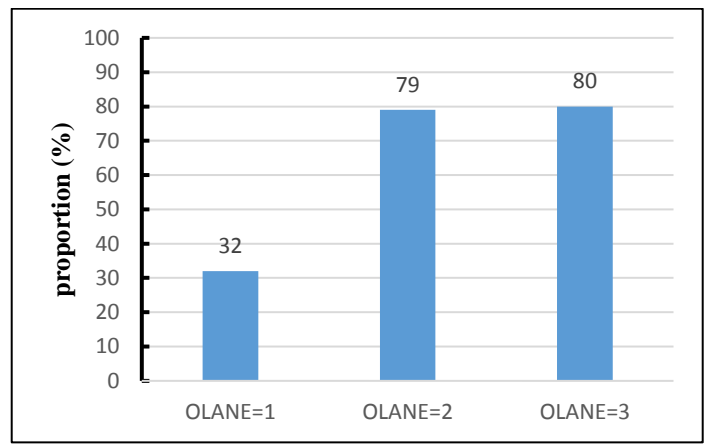

Figure 9- percentile share of accepting the first gap (LAG) by separating the lanes

## 4-9- The impact of waiting duration

Among all involved variables in pedestrian gap acceptance model, variable TIME has the smallest coefficient. However, small coefficient of this variable doesn't say that this variable doesn't affect the dependent variable because this variable is a relatively big range within 0.00s and 59s and therefore the changes of s1 for this variable aren't that much significant. According to figure. 2-I, for better investigation of TIME variable, the sensitivity of pedestrian gap acceptance probability toward the size of gap for waiting time of 10, 20 and 30 seconds was investigated. Due to this diagram, as the time of waiting increases, the possibility of pedestrian gap acceptance will decrease. For example, for gap of 3s, the possibility of accepting gap for the pedestrians who have been waiting for 10s is about 42% more than the pedestrian who has waited for 30s. the pedestrians who waits longer seem to be more cautious than other ones and as a result they look for safer gaps for crossing the street; on the contrary the pedestrians who are in more hurry, don't wait for larger gaps and cross the street in those initial moment through choosing shorter gaps.

## 5- Discussion

The results of this study show that conflict flow rate plays an important role in pedestrian decision while crossing the street. By increasing conflict flow rate, the possibility of pedestrian gap acceptance will increase and the pedestrian accepts shorter gaps. Furthermore, the pedestrians in high rate of flow are more interested in rolling movement which increases the conflict risk significantly. Similar to most of previous studies, in this study temporal gap between vehicles were studied. In some studies, however, spatial gap or both temporal and spatial gaps have been studied (Yannis et al (2013), Pawar and patil (2015). The results of this study indicates that increasing in gaps will increase the probability of gap acceptance by the pedestrian. This subject is completely consistent with the results of former studies (Sun et al, 2002; Brewer et al, 2006; Yannis et al, 2013; Cherry et al, 2012; Kadali et al, 2015 and Pawar and patil, 2016). Similar to the study of Kadali et al (2013), pedestrian gap acceptance probability will increase by drivers yielding. However, our observations show that very few of vehicles (20%) are willing to help pedestrian cross the street safely. The results also showed that by getting far from the initial point, the pedestrian gap acceptance probability will increase and he or she chooses shorter gaps to cross the street. The speed of pedestrians was another effective factors on pedestrian gap acceptance model. According to results, by increasing the speed of pedestrians, their gap acceptance probability will increase. It was also illuminated that the pedestrians adjust their speed appropriate with the size of gap and by decreasing the size of gaps, they move faster. Also in case that the pedestrian uses that very initial distance (LAG), the probability of his gap acceptance will be more. Similar to the studies of Vedagiri et al (2012), the probability of gap acceptance in those pedestrians who use rolling gap to cross the street is more. The pedestrians who have rolling movement also accept shorter gaps. According to the results of this research, group movement of pedestrians reduces the probability of gap acceptance. In another word the pedestrians who cross the street in groups, choose longer gaps. This subject is consistent with the results of studies of Yannis et al (2013) and pawar and patil (2015). Waiting duration is another effective factor on response variable. The results show that the pedestrians who wait longer are more cautious and the gap acceptance in these pedestrians is less. This subject confirms the studies of Sun et al (2002), Yannis et al (2013) and Kadali et al (2013). The type and speed of vehicles didn't have any effects on pedestrian's behavior while crossing the street. About the effect of vehicles speed on the behavior of pedestrians, similar results were formerly reported in conducted researches by Helis and Jason (1980), Parsenson et al (1996) and Yannis et al (2013). The age and gender in this research hadn't been able to define the behavior of pedestrian as well and there was no significant difference between male or female pedestrians as well as pedestrians' different age groups. Although in many studies, age and gender of pedestrians have been of the effective parameters on the behavior of pedestrians (Hamed (2001), Sun et al (2002), Oxley et al (2005)), in some studies the age and gender haven't make a significant relationship with pedestrian's behavior (Yanis (2013) and Kadali and Vedagiri (2013)).

## 6- conclusion

This research aimed to probabilistically study effective factors on pedestrian gap acceptance in uncontrolled midblock crosswalks. Variables such as individual, environmental and traffic variables were considered to define pedestrian behavior while crossing the street. A logit regression model was used in this paper in order to evaluate the probability of pedestrian gap acceptance. The results of sensitivity test showed that some parameters such as conflict flow rate, gap size, vehicles yield and the status of pedestrian while facing vehicles are of the most important factors on gap acceptance by pedestrians. The things which have been neglected in conducted probabilistic studies about pedestrian gap acceptance was the effect of traffic flow rate in the pedestrians' decision. Our results show that the characteristics such as traffic flow and gap size compared to the speed of vehicle seem to be more perceptible parameters for pedestrians. The effect of individual characteristics such as age and gender of pedestrians in their gap acceptance has been always controversial in several studies in different countries. The age and gender in this research hadn't been able to define the behavior of pedestrian. In another word, pedestrians accept gaps, regardless of their age or gender. It well shows that those variables which are related to pedestrian characteristics might function differently in different societies.